\documentclass{aa}
\usepackage{graphics}

\begin{document}

   \thesaurus{01     % A&A Section 6: Form. struct. and evolut. of stars
              (02.02.1;  % Black hole physics,
               13.07.3;  % Gamma rays: theory,
               13.07.1;  % Gamma rays: bursts,
               13.07.2)}  % Gamma rays: observations.
   \title{The Dyadosphere of Black Holes and Gamma-Ray Bursts}

%%   \subtitle{I. Overviewing the $\kappa$-mechanism}

   \author{Remo Ruffini}

%\fnmsep\thanks{Just to show the usage
%          of the elements in the author field}
%         }

   \offprints{R. Ruffini}
   \institute{I.C.R.A.-International Center for Relativistic Astrophysics
and
Physics Department, University of Rome ``La Sapienza", I-00185 Rome,
Italy\\
            email: ruffini@icra.it
%             \thanks{The university of heaven temporarily does not
%                     accept e-mails}
             }

%  \date{Received 16 May 1998/Accepted 12 July 1998}

   \maketitle

   \begin{abstract}

Recent works on the Dyadosphere are reviewed.

   \end{abstract}

%
%________________________________________________________________

\section{Introduction}

It is by now clear that  Pulsars have given the first evidence for the identification in Nature of Neutron Stars. Following the classical work of Armin Deutsch on a rotating magnetized star, from the observations of the pulsar angular velocity and its first derivative, evaluating from the theory of Neutron Stars the value of the moment of inertia, it has been possible to conclude that the energy source of Pulsars is  simply the rotational energy of Neutron Stars (Manchester \& Taylor 1977). In spite of this basic result, with the eception of a relativistic generalization of the Deutsch solution, little progress has been made in identifying the detailed mechanism of rotational energy transfer into electromagnetic energy of Pulsars.
 
The discovery of Binary X-ray sources has lead to the identification of the first Black Hole in our galaxy. The measurement of the masses of the collapsed object, made possible by the binary nature of the system, has been the discriminant between Neutron Stars and Black Holes. The matter accreting from the normal star into the deep gravitational potential well of the companion star has given evidence that the energy source of these binary X-ray sources is simply matter accretion into the deep relativistic gravitational field of a gravitationally collapsed star( Giacconi  \& Ruffini 1978). In principle more detailed analysis of the X-ray spectra and their time variability can give detailed information on the effective potential arround Black Holes (see e.g. Ruffini \&  Sigismondi 1998 and references therein), work is still today in progress.

I am proposing and give reasons that Gammma Ray Bursts for the first time we are witnessing, in real time the moment of gravitational collapse to a Black Hole. Even more important, the tremendous energies involved by the energetics of these sources, especially after the discoveries of their afterglows and their cosmological distances (Kulkarni {\it et. al.} 1998), clearly point to the necessity and give for the first time the  opportunity to use as an energy source of these objects the extractable energy of Black Holes.

That Black Holes can only be characterized by their mass-energy $E$, charge $Q$ and angular momentum $L$ has been advanced in a classical article (Ruffini \& Wheeler 1971), see Figure of
Ruffini R. \& Wheeler, J.\ A.  {\it Physics Today} 1971, ``{\it Introducing the Black Hole}'', the proof of this has been  made  after twenty five years of meticulous mathematical work. 
One of the crucial points in the Physics of Black Holes  was to realize that energies comparable to their total mass-energy could be extracted from them. The computation of the first specific example of such an energy extraction process, by a gedanken experiment, was given in (Ruffini \& Wheeler 1970) and (Floyd and R. Penrose 1971) for the rotational energy extraction from a Kerr Black hole, see Figure (2).
%                                                One column figure
%----------------------------------------------------------- fig2
\begin{figure}
\centering
\resizebox{\hsize}{8cm}{\includegraphics{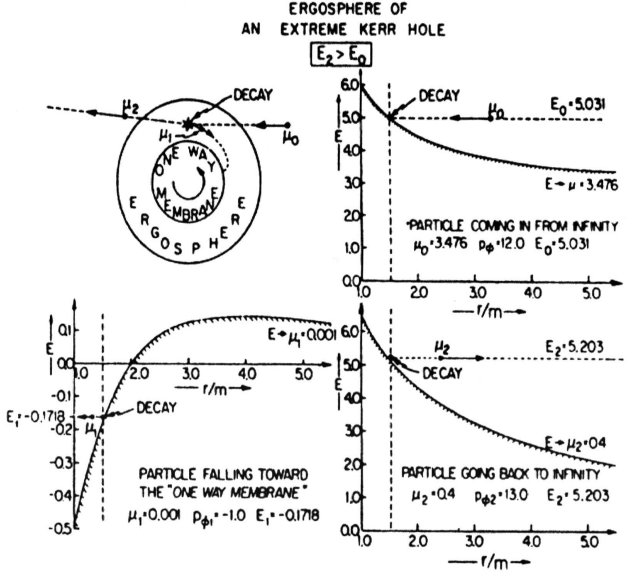}}
\caption{(Reproduced from Ruffini and Wheeler with their kind permission.) Decay of a particle of rest-plus-knetic energy $E_\circ$
into a particle which is captured into the black hole with positive
energy as judged locally, but negative energy $E_1$ as judged from      infinity, together with a particle of rest-plus-kinetic energy      $E_2>E_\circ$ which escapes to infinity. The cross-hatched curves give the effective potential (gravitational plus centrifugal) defined by the solution $E$ of Eq.(2) for constant values of $p_\phi$ and $\mu$ (figure and caption
reproduced from Christodoulou  1970).
\label{pic2}}
\end{figure}

%-------------------------------------------------------------fig2

The reason of showing this figure is not only to recall the first such explicit computation, but to emphasize how contrived and difficult such a mechanism can be:it can only work for very special parameters and should be in general associated to a reduction of the rest mass of the particle involved in the process. To slowdown the rotation of a Black Hole and to increase its horizon by the accretion of counterrotating particles is almost trivial, but to extract the rotational energy from a Black Hole, namely to slow down the Black Hole {\it and}  keep his surface area constant, is extremely difficult, as clearly pointed out also by the example in Figure (2). The above  gedanken experiments, extended as well to electromagnetic interactions,  became of paramount importance not for their direct astrophysical significance but because they gave the tool for testing the physics of Black Holes and identify their general mass-energy formula  (Christodoulou \& Ruffini 1971):
\begin{eqnarray}
E^2&=&M^2c^4=\left(M_{\rm ir}c^2 + {Q^2\over2\rho_+}\right)^2+{L^2c^2\over \rho_+^2},\label{em}\\
S&=& 4\pi \rho_+^2=4\pi(r_+^2+{L^2\over c^2M^2})=16\pi\left({G^2\over c^4}\right) M^2_{\rm ir},
\label{sa}
\end{eqnarray}
with the constraint
\begin{equation}
{1\over \rho_+^4}\left({G^2\over c^8}\right)\left( Q^4+{L^2c^2\over4}\right)\leq 1,
\label{s1}
\end{equation}
where $M_{\rm ir}$ is the irreducible mass, $r_{+}$ is the horizon radius, $\rho_+$ is the quasi spheroidal cylindrical coordinate of the horizon evaluated at the equatorial plane,
$S$ is the horizon surface area, and extreme Black Holes satisfy the equality in eq.~(\ref{s1}). The crucial point is that a transformation at constant surface area of the Black Hole, or reversible in the sense of ref.\cite{rc},  could release an energy up to 29\% of the mass-energy of an extremely rotating Black Hole and up to 50\% of the mass-energy of an extremely magnetized and charged Black Hole.

Various models have been proposed in order to tap the rotational energy of Black Holes by the processes of relativistic magnetohydrodynamic \cite{puco2},  
\cite{puco4} and references therein, see however \cite{puco1}, \cite{puco5}, though their efficiency appears to be difficult to assess at this moment. It is likely however that these of rotational energy extraction processes be relevant over the very long time scales characteristic of the accretion processes.

In the present case of the Gamma Rays Bursts a prompt mechanism, on time scales shorter then a second, depositing the entire energy in the fireball  at the moment of the triggering process of the burst, appears to be at work. For this reason we are here considering a more detailed  study of the vacuum polarization processes $à$  {\it a' la} Heisenberg-Euler-Schwinger (Heisenberg W. \& Euler H. 1931, Schwinger J. 1951)
around a Kerr-Newman Black Hole first introduced by Damour and Ruffini (Damour T. and Ruffini R., 1975).
The fundamental points  of this process can be simply summarized:

\begin{itemize}

\item They occur in an extended region arround the Black Hole, the Dyadosphere, extending from the horizon radius $r_+$ to the Dyadosphere radius $r_{ds}$ see (Preparata, Ruffini \& Xue 1998a and 1998b). Only Black Holes with a mass larger than the upper limit of a neutron star and up to a maximum mass of $6\cdot 10^{5}M_\odot$ can have a Dyadosphere, see (Preparata, Ruffini \& Xue 1998a and 1998b) for details.

\item The efficiency of transforming the mass-energy of Black Hole into particle-antiparticle pairs outside the horizon can approach 100\%, for Black Holes in the above mass range see  (Preparata, Ruffini \& Xue 1998a and 1998b) for details. 

\item The pair created are mainly positron-electron pairs and their number is much larger then the quantity $Q/e$ one would have naively expected on the ground of qualitative considerations. It is actually given by $N_{\rm pairs}={Q\over e}(1+{r_{ds}\over \hbar/mc})$, where $m$ is the electron mass.  The energy of the pairs and consequently the  emission of the associated electromagnetic radiation peaks in the X-gamma rays region, as a function of the Black Hole mass.

\end{itemize}

I shall first recall some of the results on the Dyadosphere, then consider the constituent equations leading to the expansion of the Dyadosphere

\section{ The Dyadosphere and the Energy spectrum}

We consider the collapse to amost general Black Hole endowed with an electromagnetioc field (EMBH). Following Preparata, Ruffini \& Xue (1998a and 1998b), for simplicity we consider the case of a non rotating Reissner-Nordstrom EMBH to illustrate the basic gravitational-electrodynamical process. 

%                                                One column figure
%----------------------------------------------------------- fig3
   \begin{figure}
   \resizebox{\hsize}{9cm}{\includegraphics{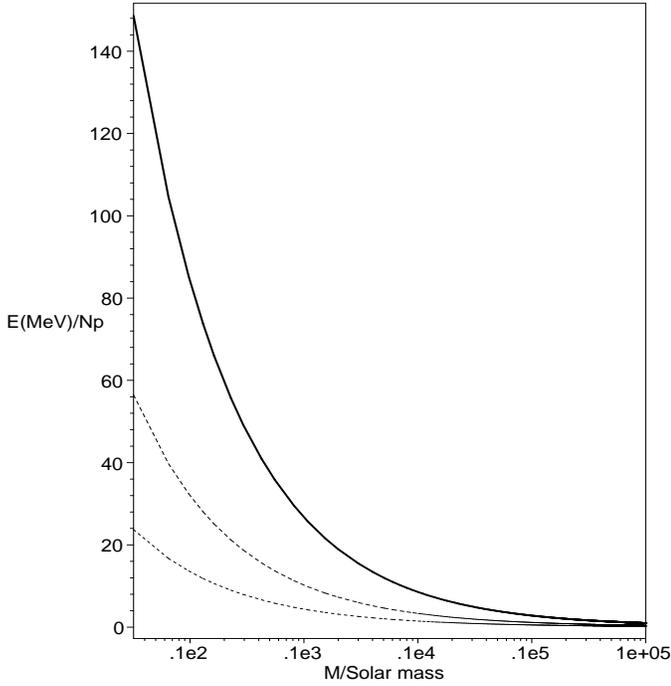}}
      \caption[]{The average energy per pair is shown here as a function of
the EMBH mass in solar mass units 
for $\xi=1$ (solid line), $\xi=0.5$ (dashed line) and
$\xi=0.1$ (dashed and dotted line) (figure and caption
reproduced from Preparata, Ruffini and Xue 1998b).
         \label{fig3}}
   \end{figure}
%
%______________________________________________________________

It is appropriate to note that even in the case of an extreme EMBH the charge-to-mass ratio is $10^{18}$ smaller than the typical charge-to-mass ratio found in nuclear matter, owing to the different strengths and ranges of the nuclear and gravitational interactions. This implies that for an EMBH to be extreme, it is enough to have a one quantum of charge present for every $10^{18}$ nucleons in the collapsing matter. 

We can evaluate the radius $r_{\rm ds}$ at which the electric field strength
reaches the critical value ${\cal E}_{\rm c}={m^2c^3\over\hbar e}$ introduced by Heisenberg and Euler, where
$m$ and $e$ are the mass and charge of the electron.  
This defines the outer radius of the Dyadosphere, which extends down to the horizon and within which the electric field strength exceeds the critical value. Using the Planck charge $q_{\rm c}= (\hbar c)^{1\over2}$ and the Planck mass $m_{\rm p}=({\hbar c\over G})^{1\over2}$, we can express this outer radius
in the form
\begin{eqnarray}
r_{\rm ds}&=&\left({\hbar\over mc}\right)^{1\over2}
\left({GM\over c^2}\right)^{1\over2} 
\left({m_{\rm p}\over m}\right)^{1\over2}
\left({e\over q_{\rm p}}\right)^{1\over2}
\left({Q\over\sqrt{G} M}\right)^{1\over2}\nonumber\\
&=&1.12\cdot 10^8\sqrt{\mu\xi} \hskip0.1cm {\rm cm},
\label{rc}
\end{eqnarray}
where
$\mu={M\over M_{\odot}}>3.2$, $\xi={Q\over Q_{\rm max}}\le 1$. 

It is important to note that the Dyadosphere radius is maximized
for the extreme case $\xi=1$ and that the region exists for EMBH's with mass larger than the upper limit for neutron stars, namely $\sim 3.2M_{\odot}$ all the way up to a maximum mass of $6\cdot 10^5M_{\odot}$.
Correspondingly smaller values of the maximum mass are obtained for values of $\xi=0.1,0.01$ as indicated in this figure. For EMBH's with mass larger than the maximum value stated above, the electromagnetic field (whose strength decreases inversely with the mass) never becomes critical. 

We turn now to the crucial issue of the number and energy densities of pairs created in the Dyadosphere. In the limit 
$r_{ds}\gg {GM\over c^2}$, we have 
\begin{equation}
N_{e^+e^-}\simeq {Q-Q_c\over e}\left[1+{
(r_{ds}-r_+)\over {\hbar\over mc}}\right].
\label{n}
\end{equation}
Their total energy is then
\begin{equation}
E^{\rm tot}_{e^+e^-}={1\over2}{Q^2\over r_+}(1-{r_+\over r_{\rm ds}})(1-
\left({r_+\over r_{\rm ds}}\right)^2).
\label{te}
\end{equation}
Due to the very large pair density 
and to the sizes of the
cross-sections for the process $e^+e^-\leftrightarrow \gamma+\gamma$, 
the system is expected to thermalize to a plasma configuration for which
\begin{equation}
N_{e^+}=N_{e^-}=N_{\gamma}=N_{\rm pair}
\label{plasma}
\end{equation}
and reach an average temperature
\begin{equation}
kT_\circ={ E^{\rm tot}_{e^+e^-}\over3N_{\rm pair}\cdot2.7},
\label{t}
\end{equation}
where $k$ is Boltzmann's constant.
The average energy per pair ${ E^{\rm tot}_{e^+e^-}\over N_{\rm pair}}$ is shown as a function
of the EMBH mass for selected values of the charge parameter $\xi$  in
Fig.3.

%                                                One column figure
%----------------------------------------------------------- fig4
   \begin{figure}
   \resizebox{\hsize}{8cm}{\includegraphics{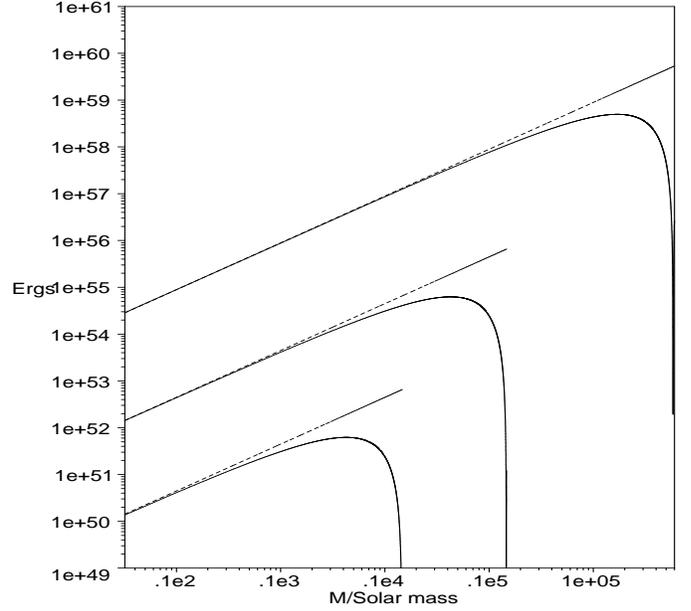}}
      \caption[]{ The energy extracted by the process of vacuum polarization is plotted (solid
lines) as a function of the mass $M$ in solar mass units for selected values of the 
charge parameter $\xi\equiv{Q\over Q_{\rm max}}=1,0.1,0.01$ for an EMBH, the case $\xi=1$ being reachable only as a limiting process. For comparison we have also plotted the maximum energy extractable from an EMBH (dotted lines) given by eq.~(\ref{em}) (figure and caption
reproduced from Preparata, Ruffini \& Xue 1998a).
\label{Fig4}}
   \end{figure}
%
%______________________________________________________________

Finally we can estimate the total energy extracted by the pair creation process in EMBH's of
different masses for selected values of the charge parameter $\xi$ and compare and contrast these values with the
maximum extractable energy given by the mass formula for Black Holes (see eqs.~(\ref{em}) and 
(\ref{s1})). This comparison is summarized in Figure (4). 
The efficiency of energy extraction by pair creation sharply decreases as the maximum value of the EMBH mass for which vacuum polarization occurs is reached. In the opposite limit the energy of pair creation processes (solid lines in Figure (4)) asymptotically reaches the entire electromagnetic energy extractable from EMBH's given by eq.(\ref{em}), leading in the limit to fully reversible transformations in the sense of \cite{rc}, $\delta M_{ir}=0$, and $100\%$ efficiency.

The discussions on the relativistic expansion of the Dyadosphere are presented in separated papers (see e.g.~Ruffini, Salmonson, Wilson \& Xue 1999).   

\section{ The relativistic expansion of the Dyadosphere and the acceleration of Cosmic Rays}

In order to insert these theoretical results in a framework suitable for making an astrophysical model for gamma ray bursts we have to analize the problem of the evolution of this plasma created in the Dyadosphere taking into due account the equation of motion of the system and the boundary conditions both at the Black hole horizon and at infinity. This problem is clearly not affordable anlitically and has been approched by us numerically both by the use of supercomputers at Livermore and by the use of a very simplified  approach in Rome. Some results are contained in the ref.\cite{rswx}.
 
	Before concluding I would like to return to the suggestion, advanced by Damour and Ruffini, that a discharged EMBH can be still extremely interesting from an energetic point of view and responsible for the acceleration of ultrahigh energy cosmic rays. I would like just to formalize this point with a few equations: It is clear that no matter what the initial conditions leading to the formation of the EMBH are, the final outcome after the tremendous expulsion of the PEM pulse will be precisely a  Kerr Newman solution with a critical value of the charge. If  the Background metric has a Killing Vector, the scalar product of the Killing vector and the generalised momentum 
\begin{equation}
P_\alpha = m U_\alpha + e A_\alpha,
\label{gm}
\end{equation}
is a constant along the trajectory of any charged gravitating particle following the relativistic equation of motion in the background metric and electromagnetic field (Jantzen and Ruffini 1999). Consequentely electron (positron) starting at rest in the Dyadosphere will reach infinity with an energy $E_{kinetic}\sim 2mc^2({GM\over c^2})/({\hbar\over mc})\sim 10^{22}$eV for $M=10M_\odot$.

\end{document}